\documentclass[aps,prb,twocolumn,amsmath,amssymb,showpacs]{revtex4-1}
\usepackage{graphicx}
\usepackage{dcolumn}
\usepackage{bm}
\begin{document}

\title{Spin-Charge Locking and Tunneling into a Helical Metal}

\author{P. Schwab$^{a}$, R. Raimondi$^b$, C. Gorini$^c$}
\affiliation{$^a$Institut f\"ur Physik, Universit\"at Augsburg, 86135 Augsburg, Germany\\
$^b$CNISM and Dipartimento  di Fisica "E. Amaldi", Universit\`a  Roma Tre, 00146 Roma, Italy\\
$^c$Institut de Physique et Chimie des Mat\'eriaux de Strasbourg (UMR 7504),
CNRS and Universit\'e de Strasbourg, 23 rue du Loess, BP 43, F-67034 Strasbourg Cedex 2, France
}

\date{\today}

\begin{abstract}
We derive a kinetic equation for the electrons 
moving on the surface of a three-dimensional topological insulator.
Due to the helical nature of the excitations
backward scattering is suppressed in the collision integral,
and the spin dynamics is entirely constrained by that of the charge.
We further analyze 
the tunneling between the helical and a conventional metal or ferromagnet.
We find that the tunnel resistance strongly depends on the angle between the
magnetization in the ferromagnet and the current in the helical metal.
A nonmagnetic layer on top of the helical metal amplifies the current-induced spin polarization.
\end{abstract}

\maketitle

Topological insulators\cite{moore2009,hasan2010} have recently attracted considerable interest,
especially after their experimental discovery in two-\cite{konig2007} and three dimensions
\cite{hsieh2009,chen2009,xia2009,kuroda2010}.
While insulating in the bulk, such materials possess gapless helical
edge states whose existence depends on -- and is protected by -- time reversal invariance
\cite{kane2005, kane2005a,bernevig2006,moore2007,fu2007,murakami2007,roy2009}.
This makes the latter robust against time-reversal symmetric perturbations
(such as impurity scattering) and at the same time very sensitive to 
time-reversal breaking ones (such as magnetic fields).
When the topological insulator is a three-dimensional system, the gapless excitations are confined 
to its surface and form a two-dimensional conductor
and presents novel and interesting properties, see for example
Ref. [\onlinecite{culcer2010}] for a recent summary.
In particular, Burkov and Hawthorn\cite{burkov2010} considered the problem of
spin-charge coupled transport on a helical metal and derived diffusion equations for charge and spin.
They predicted a distinctive magnetoresistance effect 
when the helical metal is placed between a ferromagnet and a normal metal. 
In this paper we extend their work in several ways. We first derive a
kinetic equation which is valid even beyond the diffusive regime. In this latter regime we obtain a 
diffusion equation which agrees with that of Burkov
and Hawthorn as far as the charge component is concerned. 
On the other hand for the spin density we find a different
behaviour, namely the spin dynamics is constrained to follow the charge one.
Secondly, we consider the effect of bringing the helical metal in
contact with a ferromagnet and discuss its unconventional
magnetoresistance.

For the simplest case the effective Hamiltonian describing the surface states of a
topological insulator has the form
\cite{zhang2009,hasan2010}
\begin{equation}
\label{eq1}
H=v_F {\bf k}\times { {\bf e }}_z\cdot {\boldsymbol \sigma},
\end{equation}
where the parameter $v_F$ is the velocity of the gapless excitations, 
${ { \bf e}}_z$ is a unit vector perpendicular to the surface, 
${\bf k}$ is the two-dimensional momentum operator, ${\boldsymbol\sigma}$ 
are the Pauli matrices, and units of measure such that $\hbar=1$ have been used.
The eigenstates of $H$ form two bands with linear dispersion,
$\epsilon_\pm = \pm v_F k$, and we will assume in the following that
the Fermi energy is located deep enough in the upper band for states
in the lower band to remain fully occupied and thus
not relevant for the dynamics of the system.

Because of the helical nature of the excitations, such a surface conductor
is called a helical metal and presents novel and interesting
properties. For instance the velocity operator is given by $  \dot {\bf x} = v_F {\bf e}_z \times
{\boldsymbol \sigma }$, so that the particle current becomes
${\bf j} = 2 v_F  {\bf e}_z \times {\bf s}$, ${\bf s}$ being the
spin polarization. This means that the particle current is entirely
constrained by the spin density or that, vice versa, the in-plane components of the spin
density are constrained by the particle current.

Such a constraint will also become apparent later in the kinetic equation 
for a disordered helical metal. The equation will be 
valid when the Fermi energy is far from the Dirac point and satisfies the condition
$\epsilon_F \gg 1/\tau$, with $\tau$ the scattering time.
We will follow a procedure similar to what was done for graphene in Ref.~[\onlinecite{kechedzhi2008}].
The starting point is the retarded Green function which in the absence
of disorder reads
\begin{equation}
\label{eq2}
G^R_{ss'}=G^R_0[\sigma_0]_{ss'}+{\bf G}^R\cdot [{\boldsymbol
\sigma}]_{ss'},
\end{equation}
where
\begin{equation}
\label{eq3}
G_0^R =\frac{1}{2}\left( G^R_++G^R_-\right)
\end{equation}
and
\begin{equation}
\label{eq4}
{\bf G}^R= 
\frac{1}{2}
{\hat {\bf k}}\times {\bf e}_z 
\left( G^R_+-G^R_-\right),
\end{equation}
with
\begin{equation}
\label{eq5}
G_{\pm}^{R}=\left(\epsilon \mp v_F k+\mu +{\rm i}0^+\right)^{-1}
\end{equation}
and $\hat {\bf k } $ being the unit vector in the ${\bf k}$-direction.
For clarity we included in Eq.~(\ref{eq2}) the spin indices $s$ and
$s'$.
The Green function ${\check G}$ has the two-by-two matrix structure of the Keldysh
formalism
\begin{equation}
\check G ({\bf R}, T; {\bf k}, \epsilon )=
\begin{pmatrix}
G^R({\bf R}, T; {\bf k}, \epsilon )      &  G ({\bf R}, T; {\bf k}, \epsilon ) \\
   0   &  G^A({\bf R}, T; {\bf k}, \epsilon )
\end{pmatrix}
,\end{equation}
where ${\bf R}$ and ${T}$ are the center-of-mass space and time coordinates, 
while ${\bf k}$ and $\epsilon$ are the Fourier transformed variables of the relative coordinates.
The left-right subtracted Dyson equation reads
\begin{equation}
\label{eq8}
\partial_T  \check G +  \frac{v_F }{2}  \left\{ {\bf e}_z \times {\boldsymbol
\sigma}\cdot \partial_{\bf R}, \check  G\right\}
+{\rm i}\left[ H, \check G\right]=-{\rm i}\left[ \check \Sigma,  \check
G\right],
\end{equation}
where $[, ]$ and $\{ , \}$ are the commutator and anticommutator.
On the right-hand-side of the equation the self-energy $\check \Sigma$ appears. 
For a delta-correlated impurity potential with
$\langle V({\bf x}) V({\bf x}') \rangle  =u^2 \delta({\bf x}-{\bf x}')
$ and within the Born approximation the self-energy is proportional to
the Green function, integrated over the momentum
\begin{equation}
\check \Sigma({\bf R}, T; \epsilon ) = u^2  \int \frac{d^2 k}{(2 \pi)^2} \,  \check G({\bf R}, T; {\bf k},
\epsilon ).
\end{equation}
We define a quasiclassical Green function as 
\begin{equation}
\label{eq6}
{\check g}({\bf R}, T; {\bf {\hat k}}, \epsilon )= 
   \frac{{\rm i}}{\pi}\int {\rm d}\xi {\check G}({\bf R}, T; {\bf { k}}, \epsilon ),
\end{equation}
where $\xi = v_F k - \mu$ and the integration is performed in the
vicinity of the Fermi energy.
From Eqs.~(\ref{eq2})-(\ref{eq5}), one obtains the retarded component of
$\check g$ for energies close to the Fermi level ($|\epsilon | \ll v_F k_F $) as
\begin{equation}
\label{eq7}
g^R=\frac{1}{2}+ \frac{1}{2}{\hat {\bf k}}\times {\bf e}_z\cdot {\boldsymbol \sigma}
,\end{equation}
i.e. $g^R$ is a projector on the upper band of the
Hamiltonian (\ref{eq1}).
The self-energy reads
\begin{equation}
\check  \Sigma = -\frac{ \rm i}{ \tau } \langle \check g \rangle  
,\end{equation}
where $1/\tau=\pi N_0 u^2$, $N_0=k_F /(2\pi v_F)$ is the single-particle density of states 
at the Fermi energy
and $\langle \check g \rangle$ is the average of the quasiclassical
Green function over the Fermi surface.
Finally, for the Keldysh component of the Green function we find from Eq.~(\ref{eq8}) the
kinetic equation 
\begin{eqnarray}
\lefteqn{ \partial_T g + \frac{v_F}{2} \left\{ {\bf e}_z \times {\boldsymbol \sigma}\cdot \partial_{\bf R}, g\right\}
 + {\rm i} v_F k_F \left[ {\hat {\bf k}}\times {\bf e}_z \cdot
{\boldsymbol \sigma}, g\right] } &&
\nonumber\\
  & = &  -  \frac{1}{\tau}g+\frac{1}{\tau}\langle g\rangle 
  +\frac{1}{2\tau}  \left\{ {\hat {\bf k}}\times {\bf e}_z \cdot  {\boldsymbol \sigma}, \langle g\rangle \right\} \label{eq9}
.\end{eqnarray}
Notice that this is a matrix equation in spin space, 
$g_{ss'} = g_0 [\sigma_0]_{ss'} + {\bf g} \cdot [ {\boldsymbol
\sigma}]_{ss'}$.  However, its structure can be considerably simplified.
Equation (\ref{eq9}) is derived under the assumption that $v_F k_F$ is much larger than all other energy
scales in the problem.  Therefore, the leading contribution to $g$ must
commute with ${ \hat {\bf k} }\times {\bf e}_z\cdot {\boldsymbol
\sigma}$, or in other words, $g$ is diagonal in the eigenstates of $H$. 
Since the lower band has no density of states at the Fermi
level, only the upper band contributes to $g$ which is then
proportional to the upper-band projector, i.e.
\begin{equation}
\label{eq13}
g_{ss'} = g_0 \left(  [\sigma_0]_{ss'} + \hat {\bf k} \times {\bf e}_z
\cdot [ {\boldsymbol \sigma}]_{ss'} \right)
.\end{equation}
From the spin-trace of Eq.~(\ref{eq9}) we obtain then a kinetic equation
for $g_0$,
\begin{equation} \label{eq14}
\partial_T g_0 + v_F \hat {\bf k } \cdot \partial_{\bf R} g_0 = - \int
\frac{d \varphi'}{2 \pi } W(\varphi-\varphi')[ g_0(\varphi) -
g_0(\varphi') ]
,\end{equation}
which is just the standard kinetic equation for a metal with
an angle dependent scattering potential \cite{rammer1986,schwab2003}.
Here, starting from a short range potential, we find
\begin{equation} 
W(\varphi, \varphi') = \frac{1}{\tau}  
(1 + \hat {\bf k } \cdot  \hat {\bf k}') =
\frac{1}{\tau} ( 1 + \cos(\varphi -
\varphi')  )
,\end{equation} 
where the cosine term accounts for the absence of
backscattering.
The spin-dependent contributions to $g$ can be reconstructed from
$g_0$, cf. Eq.(\ref{eq13}): 
\begin{equation}g_x  \approx \hat k_y  g_0 ,  \mbox{ and } g_y \approx -
\hat k_x g_0  \label{eq16}
.\end{equation}
The $g_z$ component is  nonzero only to 
subleading order in $1/v_F k_F$ and is after some algebra determined as
\begin{equation}
g_z  \approx \frac{1}{v_F k_F} \left( 
  \frac{\hat k_x}{\tau} \langle \hat k_y g_0 \rangle 
- \frac{\hat k_y}{\tau} \langle \hat k_x g_0 \rangle + v_F \hat k_y
  \partial_x g_0 - v_F \hat k_x  \partial_y g_0
\right). \label{eq17}
\end{equation}
Equations (\ref{eq14}), (\ref{eq16}) and
(\ref{eq17}) are one of the main results obtained in this paper. 

It is instructive to study the angular average of the kinetic
equation, since the latter is closely related to the continuity
equation for the observables.
Recall, for example, that the particle and spin densities are obtained 
from the quasiclassical Green function by taking the angle average
and integrating over the energy \cite{rammer1986},
\begin{equation}
\label{tun_6}
n =  - \frac{N_0}{2} \int {\rm d}\epsilon  \langle g_0\rangle + N_0 e
\phi,  \
{\bf s} = - \frac{N_0}{4} \int {\rm d} \epsilon \langle {\bf g }
\rangle
,\end{equation}
where $\phi$ is the scalar electrical potential.
From Eq. (\ref{eq14}) we obtain the continuity equation for the
density as
\begin{equation}
 \partial_T \langle g_0 \rangle + 
  \partial_{\bf R} \cdot  \langle v_F \hat {\bf k}  g_0 \rangle = 0
.\end{equation}
Using Eq.~(\ref{eq16}) we find $v_F \langle \hat k_x  g_0 \rangle =
-v_F \langle g_y \rangle$ and $v_F \langle \hat k_y g_0 \rangle = v_F
\langle g_x \rangle$,
i.e. we verify the general relations between the particle current and
the in-plane spin density stated already below Eq.~(\ref{eq1}).
For the spin density we find 
\begin{eqnarray} \label{eq20}
\partial_T s_x  +\frac{v_F}{4} \partial_y n +
\frac{1}{2\tau} s_x  &= &0 \\
\label{eq21}
\partial_T s_y  -\frac{v_F}{4} \partial_x n + \frac{1}{2\tau} s_y  &= &0 
,\end{eqnarray}
from which we identify $2 \tau $ as the spin relaxation time.
In (\ref{eq20}) and (\ref{eq21}) we ignored the scalar electric
potential, $\phi$.

In order to compare with Ref. [\onlinecite{burkov2010}] we will now
discuss the diffusive limit. 
The particle density obeys the diffusion equation
\begin{equation}
\label{per_20}
\partial_T n - D \partial_{\bf R}^2 n  = 0 , \ D=
\frac{v_F^2}{2}\tau_{tr}, \tau_{tr} = 2 \tau
,\end{equation}
i.e., the particle current is
${\bf j }= - D \partial_{\bf R} n $. 
The transport time  $\tau_{tr}$ being twice as long as the scattering
time $\tau$ stems from the absence of
backscattering in the helical metal.
Equation (\ref{per_20}) is consistent with Ref.
[\onlinecite{burkov2010}], notice however the different definition of the diffusion constant.
Since the spin relaxation time is very short (it equals the transport
scattering time)
the spin dynamics is not diffusive. However, Eqs. (\ref{eq20}) and
(\ref{eq21}) are still valid. In the diffusive limit the time
derivative of the spin-density is small compared to the  spin
relaxation term, i.e. the spin density is given by the spatial
derivative of the charge density so again we identify the general
relation between particle current and spin density. 

We will now analyze the transport through the helical metal when it is
contacted via a tunnel junction to a ferromagnet.
We will find an unconventional magnetoresistance effect that arises
since the tunneling probability between the helical metal and
the ferromagnet is strongly angle dependent:
the overlap of two spinors with polarization in $\hat {\bf m
}$- and   $\hat {\bf m}'$-directions depends on the angle between the
two vectors,
$| \langle \hat { \bf m } | \hat {\bf m}' \rangle |^2 = \frac{1}{2} +
\frac{1}{2} \hat { \bf m } \cdot  \hat { \bf m }'  $.
The states in the conduction band of the helical metal are polarized as
$\hat {\bf m}' =  \hat  { \bf k }  \times {\bf e}_z$,  so
that the tunneling probability from a state $\hat {\bf m}$ in the
ferromagnet into a state in the 
helical metal with momentum ${\bf k}$ depends on the angle between $\hat {\bf m}$ and ${\bf k}$.

In order to make these considerations more formal we introduce the tunneling Hamiltonian
\begin{equation}
\label{tun_1}
H_t=\sum_{s}\int {d^2 x} \,  t({\bf x})  \psi^{\dagger}_s ({\bf x})
\psi_{F,s} ({\bf x}) + c.c.,
\end{equation}
where $\psi^{\dagger}_s ({\bf x})$ is the field operator for an
electron with spin $s$ in the
helical metal and $\psi_{F,s} ({\bf x})$ that for the ferromagnet. 
For a point-like (on the quasiclassical scale) contact at ${\bf x}=0$
the tunneling amplitude is
\begin{equation}
\label{tun_2}
t({\bf x}) = t \delta ({\bf x}).
\end{equation}
Eq.~(\ref{tun_1}) leads then to an additional contribution to the self-energy of the form
\begin{equation}
\label{tun_3}
\Sigma_t({\bf R}, T; \epsilon ) = |t|^2\delta({\bf R})   \int
\frac{d^2
k}{(2\pi)^2} \check G_F({\bf R}, T; {\bf k}, \epsilon)
\end{equation}
where 
$\check G_F$ is  the Green function  of the ferromagnet. 
We assume that the conduction electrons in the ferromagnet can be described in terms of an
incoherent superposition of majority and minority carriers, so  
we write
\begin{equation}
\label{tun_4}
\check G_F= \check  G_{F,\uparrow}P_{\uparrow}+ \check G_{F,\downarrow}P_{\downarrow}, 
\end{equation}
where $P_{\uparrow,\downarrow}=(\sigma_0 \pm {  \hat {\bf  m} }\cdot
{\boldsymbol\sigma})/2$ projects on states parallel
antiparallel to 
${ \hat {\bf   m} }$.
The kinetic equation (\ref{eq14}) becomes
\begin{eqnarray}
(\partial_T &+&v_F {\hat {\bf k} } \cdot \partial_{\bf R})g_0  =  -\frac{1}{\tau}g_0
      +\frac{1}{\tau}\left( \langle g_0\rangle+{\hat {\bf  k} } \cdot
      \langle {\hat {\bf k}' }g_0\rangle \right)\nonumber \\
 &-& 
 \pi |t|^2  \delta ({\bf R})N_{\uparrow}
    (1+{ \hat {\bf m} }\cdot {\hat {\bf  k} }\times {\hat {\bf e }}_z)
    ( g_0 -\frac{1}{2}\langle g_{\uparrow}\rangle) \nonumber \\
 &-&   
 \pi |t|^2  \delta ({\bf R})N_{\downarrow}
     (1 -{ \hat {\bf  m} }\cdot {\hat{\bf k}}\times {\hat {\bf e}}_z)
     ( g_0  -\frac{1}{2}\langle g_{\downarrow}\rangle),\label{tun_5}
\end{eqnarray}
where $g_{\uparrow, \downarrow}$ and $N_{\uparrow ,\downarrow}$ are the
quasiclassical Green function and the density of states in the
ferromagnet. 
The terms in the second and third line of this equation
describe tunneling between the helical metal and the spin up or down
band of the ferromagnet.
As anticipated above, the tunneling probability between the helical
metal and the ferromagnet is a $\hat {\bf k}$-dependent function.
After the angular average and integrating over the energy we obtain the following continuity equation for the charge
density,
$\rho = (-e ) n$,
\begin{eqnarray}
\partial_T \rho +\partial_{\bf R}\cdot {\bf j} 
&=& 
      - \delta({\bf R})\left[
      G_{\uparrow}(U-U_{\uparrow})+ G_{\downarrow}(U-U_{\downarrow})\right]
\nonumber\\
&-& 
\delta ({\bf R})
 \frac{G_{\uparrow}-G_{\downarrow}}{e^2N_0 v_F} {\hat {\bf e}}_z \times {\hat {\bf  m} }\cdot {\bf j},
 \label{tun_7}
\end{eqnarray}
having introduced the tunneling conductances 
\begin{equation}
\label{tun_8}
G_{\uparrow,\downarrow}=  \pi e^2 |t|^2 N_0 N_{\uparrow,\downarrow}
\end{equation}
and the integrals
\begin{equation}
\label{tun_9}
U =\frac{1}{2e}\int {\rm d}\epsilon \langle g_0\rangle, \
U_{\uparrow,\downarrow}=\frac{1}{4e}\int {\rm d}\epsilon \langle g_{\uparrow\downarrow}\rangle.
\end{equation}
The latter have the meaning of a voltage, or more precisely of a
magneto-electrochemical potential\cite{johnson1987}, cf. Eq. (\ref{tun_6}). 
Sources and sinks for the charge density in the helical metal appear
on the right hand side of the equation due to tunneling.
One observes that a tunnel current cannot only be driven by a voltage
across the contact, but also by a current bias in the helical metal.

\begin{figure}
\includegraphics[width=0.45\textwidth]{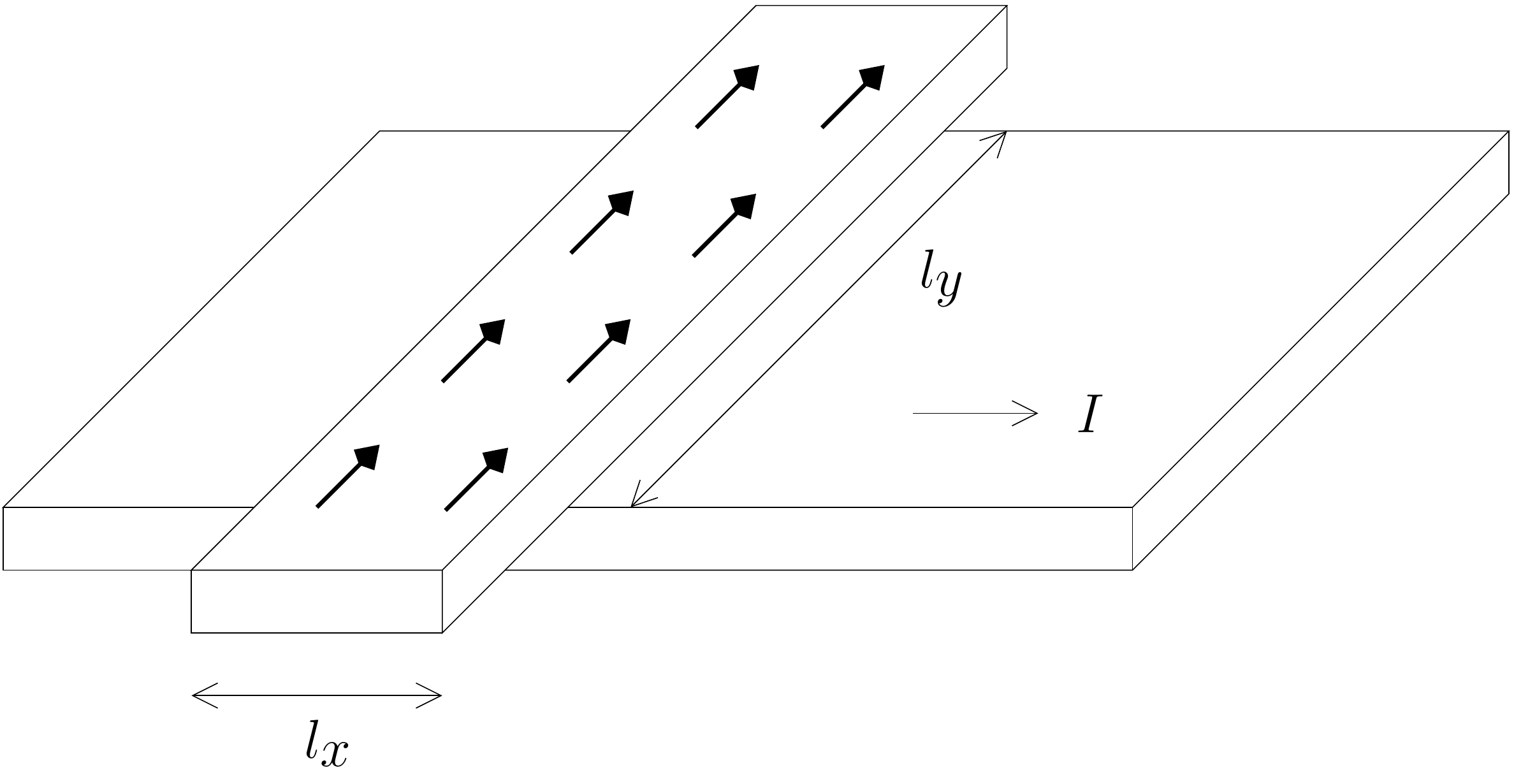}
\caption{  \label{fig1}
Schematic view of system under consideration: electrons can tunnel between
a ferromagnet (top) and a helical metal (bottom). The size of the
contact is $l_x$ times $l_y$, the current is assumed to flow only in
$x$-direction.
}
\end{figure}

For simplicity we will assume in the following equal magneto-electrochemical
potentials for spin up and down electrons in the ferromagnet,
$U_{\uparrow\downarrow} = U_\uparrow = U_\downarrow$. 
Furthermore, instead of a point-like tunnel contact we will consider
an extented tunneling region as depicted in Fig. \ref{fig1}. 
This is achieved by replacing the
$\delta$-function on the right hand side of Eq. (\ref{tun_7})
by an appropriate function characterising the shape of the contact.
In particular, we will study in detail
a junction with a long extension $l_y$ in
$y$-direction
\footnote{Our equations apply to wide contacts provided they can be modeled as a
series of incoherent tunnel junctions}.
We replace then the $\delta$-function in Eq.~(\ref{tun_7}) by
\begin{equation}
\delta({\bf R} )  
 \rightarrow \Bigg\{
    \begin{array}{cl}
        l_y^{-1} \delta(x)    &  {\rm for  }  -l_y/2<y<l_y/2\\
                  0 &   {\rm otherwise}
     \end{array},
\end{equation}
and we assume that current flows only in $x$-direction. The
continuity equation thus becomes one-dimensional.
Due to the remaining one-dimensional $\delta$-function the current density jumps at $x=0$. The
size of this jump is determined by
integrating the continuity equation with respect to the $x$-coordinate for a
region close to the tunneling contact with the result
\begin{equation}
j_+ - j_- = - \frac{G_\uparrow + G_\downarrow }{l_y} (U -U_{\uparrow \downarrow} )
            + \frac{G_\uparrow - G_\downarrow}{l_y N_0 v_F e^2}
            \frac{j_+ +  j_-}{2}\hat m_y
,\end{equation}
where $j_\pm = j(x= \pm0)$ is the current right and left from the
tunnel contact and we assumed that 
$\delta(x) j(x) = \delta (x) (j_+ + j_-)/2$.
When we fix the current left to the contact to zero we determine the
tunnel resistance as
\begin{equation}
\label{tunnel_resistance}
R_t = \frac{1}{G_\uparrow + G_\downarrow} - 
\frac{\hbar}{e^2}\frac{\lambda_F \hat m_y }{2l_y}  \frac{ G_\uparrow -
G_\downarrow}{ G_\uparrow + G_\downarrow} 
,\end{equation}
where for clarity we put back $\hbar$.
Through $\hat m_y$ the tunnel resistance  depends on the orientation of
the ferromagnet with respect the the $x$-axis, similar to what was
found by Burkov and Hawthorn \cite{burkov2010}.
The relative size of the
magnetoresistance effect increases with increasing tunnel conductance being controlled by the
dimensionless parameters $(G_\uparrow -G_\downarrow)/(e^2 /\hbar )$ and 
$\lambda_F /l_y $.
Apparently the tunnel resistance may even become negative for large enough values of these two parameters.
However, as we will see below, when this happens equation (\ref{tunnel_resistance}) 
is no longer valid and a more careful treatment which takes into
account a finite contact area of the tunnel junction is needed. 
To analyze a junction with a finite width $l_y$ and length  $l_x$  we make the following
 replacement in Eq.~(\ref{tun_7}):
 $\delta ({\bf R}) \to (l_xl_y)^{-1}$ when ${\bf R}$ is inside the tunnel
 junction and $\delta({\bf R}) \to 0$ otherwise.
This leads to the tunnel resistance
\begin{equation}
\label{new_tunnel}
R_t= \frac{1}{G_\uparrow + G_\downarrow} 
       f\left(\frac{\lambda_F}{l_y}\frac{G_\uparrow -G_\downarrow}{e^2 /\hbar}  \hat m_y\right),
\end{equation}
with the function $f(x)=x/(e^x-1)$. 
For small $x$, $f(x)\approx 1-x/2$ and one recovers Eq.~(\ref{tunnel_resistance}).
When $x$ increases and becomes of order one, the resistance remains positive, 
but depends strongly on the sign of $x$, i.e., the orientation of the
ferromagnet.
For $x>0$, the resistance decreases exponentially 
in the parameter $(G_\uparrow -G_\downarrow)/(e^2 /\hbar )(\lambda_F /l_y) $, 
while it increases linearly for  $x < 0$.
The tunnel junction is then acting as a spin-diode.

In this final part of the paper we consider
tunneling from the helical metal into a normal, nonmagnetic metal.
The tunnel resistance for the charge becomes $R_t = 1/(G_\uparrow +
G_\downarrow)$ since spin up and down have both
the same density of states, $N_\uparrow = N_\downarrow = N_{\uparrow \downarrow}$ and
also identical tunnel conductance. Nevertheless tunneling into a normal metal
is of interest since, as we will demonstrate, the helical metal
injects spin into the normal metal.
Going through the same steps as in Eqs.~(\ref{tun_3})--(\ref{tun_7})
but now for the spin-density in the normal metal we arrive at
\begin{eqnarray}
\partial_t s^a + \partial_{\bf R} \cdot {\bf j}^a & =  &-
   \frac{\hbar/2e^2}{R_t} 
   \frac{1}{N_{\uparrow\downarrow}} \delta({\bf R}) s^a  \nonumber\\[1ex]
&& - \frac{\hbar/2e^2}{R_t} \frac{\delta({\bf R})}{e N_{0} v_F} 
({\bf j } \times {\bf e}_z)_a
,\end{eqnarray}
where ${\bf j}^a$ is the spin current in the normal metal and ${\bf j
}$ the charge current in the helical metal. Whereas the tunneling of
charge is controlled by the voltage across the tunnel junction, the tunneling of spin
is controlled by the current density in the helical metal, i.e.
a current flowing parallel to the junction. Thus
injection of a pure spin current into the normal metal is possible.
Furthermore, if spin tunneling in and out are balanced  so that there is no spin current injection,
there is a steady state determined by 
\begin{equation}
\frac{1}{N_{\uparrow \downarrow}}{\bf s} = - \frac{1}{e N_0 v_F}  {\bf j } \times {\bf e}_z
,\end{equation}
which is equivalent to 
\begin{equation}
{\bf s}\Big|_{\text{ normal metal }} = \frac{2 N_{\uparrow \downarrow}}{N_0}
{\bf s}\Big|_{\text{ helical metal}}
.\end{equation}
A normal metal on top of the helical metal thus amplifies the current-induced spin polarization.

We finally turn to the question of how robust the effects found are, having in mind that the effective
Hamiltonian (\ref{eq1}) is only valid in the vicinity of the Dirac
point, whereas the interpretation of the experimentally observed Fermi
surfaces \cite{hsieh2009,chen2009,kuroda2010} requires at finite doping terms that are quadratic and even cubic
in ${\bf k}$ \cite{fu2009}. For example, to third order
in ${\bf k}$, the effective
Hamiltonian for the surface states of the topological insulators
Bi$_2$Te$_3$ and Bi$_2$Se$_2$ is \cite{fu2009}
\begin{equation} \label{eq35}
H = 
v {\bf k } \times {\bf e}_z \cdot \boldsymbol \sigma + 
\frac{k^2}{2m^*} + 
\frac{\lambda}{2}(k_+^3 + k_-^3) \sigma_z
,\end{equation}
where the Dirac velocity contains a second order correction, $v= v_F(1+ \alpha k^2)$, 
$k_\pm = k_x \pm i k_y $, and $\alpha$, $m^*$,$\lambda$ are the parameters characterizing the strength of
the higher order corrections.
In order to understand the effect these extra terms in the Hamiltonian
have on the magnetoresistance and the spin injection, recall that
the origin of both is the helicity of the conduction electrons.
The quadratic term in the Hamiltonian (\ref{eq35}) has no spin
structure, does not affect the helicity of the eigenstates, and
therefore cannot qualitatively change our results.
The cubic term on the other hand disturbs the helicity of the
eigenstates (the angle between the velocity and the spin now
depends on the position on the Fermi surface), so for a strong cubic
term we leave the region where our results are reliable. 

In conclusion we have derived a kinetic equation for 
a helical model and have shown that the spin dynamics is constrained to follow that of the charge. 
In the diffusive regime for charge, the spin density is described in terms
of the charge density gradient. Furthermore, when the helical metal is
placed in contact with a ferromagnetic metal, the tunneling current  
depends on the relative orientation between the current and the  polarization 
in the ferromagnet. In the limit of large tunneling conductance, the device acts as a spin-diode.
A helical metal in contact with a normal metal injects spin into the
latter, with a rate that can be controlled independently from the
injection of charge carriers. 

We thank the Deutsche Forschungsgemeinschaft (SPP1285)
and the French Agence Nationale de la Recherche (grant no. ANR-08-BLAN-0030-02) for financial support.
\bibliography{paper_hm}
\end{document}